\documentclass[seceq,supplement]{ptptex}

\usepackage[xdvi,dvips]{graphicx}

\newcommand{\btem}{\bibitem}
\newcommand{\beq}{\begin{eqnarray}}
\newcommand{\eeq}{\end{eqnarray}}

\newcommand{\la}{\langle}
\newcommand{\ra}{\rangle}

\newcommand{\dirac}{\partial\llap{$\diagup$\kern-2pt}}
\newcommand{\kr}{K'/K}

\title{
Axial Anomaly, Mismatched Fermi Surfaces and Vector Interaction
in Dense Neutral Quark Matter%
}

\author{
Teiji \textsc{Kunihiro}$^{1}$%
and
Zhao \textsc{Zhang}$^{2}$%
}

\inst{
  ${}^1$Department of Physics, Kyoto University,
606-8502, Kyoto, Japan
\\
  ${}^2$School of Mathematics and Physics, North China Electric
Power University, Beijing 102206, China
}

\abst{
We report on effects of the q-$\bar{\rm q}$ vector interaction
and/or the U(1)$_A$-anomaly-induced chiral-diquark coupling
on the charge-neutral quark matter in $\beta$-equilibrium.
We show that when the vector coupling is absent,
there can appear a cross-over region sandwiched by two critical points
in the intermediate temperature ($T$) region,
while the phase transition
in the low-$T$ region including zero temperature keeps being first order
until the strength of the anomaly term is increased to have a critical value.
On the other hand, when the vector coupling is also present,
there appears a crossover region in the low-$T$ area including
zero temperature with a new critical point, as was first
demonstrated by Kitazawa et al and the present authors without and with
the charge-neutral condition, respectively.
We remark that the possible chromomagnetic instability
 is suppressed and can be even completely absent
 owing to the enhanced diquark coupling due to the anomaly term and/or by the vector interaction.
}

\begin{document}

\maketitle

\section{Introduction}
In this report, we 
discuss the phase diagram
of the charge-neutral
quark matter under beta-equilibrium constraint,
taking into account 
 the  q-$\bar{\rm q}$ vector interaction
and/or U(1)$_A$-anomaly-induced chiral-diquark coupling\cite{arXiv:1102.3263}.

\subsection{Effect of the vector-type interaction in chiral transition}

The significance of the vector-vector interaction $\sim G_V(\bar{q}\gamma^{\mu}q)^2$
on the chiral phase transition in hot and dense quark matter
 is known for a long time, and clearly demonstrated by
 Kitazawa et al \cite{hep-ph/0207255}, who showed
 that the increase of the vector coupling $G_V$ enlarges the crossover region,
 and the celebrated QCD critical point\cite{288973} eventually disappears
completely in the phase diagram with a sufficiently large coupling:
See Fig. 13 of \citen{hep-ph/0207255}.

When the possible color-superconducting(CSC) phase transition is
taken into consideration,
 the phase boundary for the chiral-to-CSC phase transition
becomes of crossover in the low temperature ($T$) region including vanishing temperature,
 as is shown in Fig.8 of \citen{hep-ph/0207255} by Kitazawa et al.

\begin{figure}[thb]
\begin{center}
{\includegraphics[scale=.65]{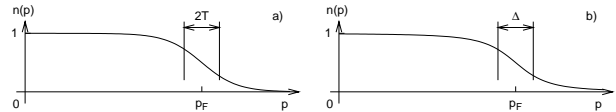}}
\caption{
The schematic figures to show that the diquark
gap plays a similar role as temperature for the chiral transition:
(a) the distribution function at finite $T$, $n = 1/(e^{(p-\mu)/T}+1)$
and (b) that for CSC phase with a diquark gap $\Delta$. Taken from
\citen{addenda}.}
\end{center}
\label{fig:analogy}
\end{figure}

The reason why the phase transition involving the chiral restoration becomes so weak
and turns to  crossover is a smearing of the Fermi surface due to the diquark gap,
 which is analogous to
that due to finite $T$  where the quark distribution function is smeared
and  the chiral transition is known to become crossover at zero chemical potential:
See Fig.~1.
Notice that a positive energy state of the fermion has a positive matrix element
 $\la \bar{q}q \ra$ while the vacuum chiral
condensate has a negative value,
 and hence the presence of the positive-energy fermions act to decrease the chiral condensate in the
absolute value.

Such a competition between the chiral and CSC phase transition is enhanced by the
vector interaction because the repulsive term $G_V\rho^2$
 postpones the emergence of the high density, although the large Fermi sphere is preferable
for the formation of the CSC phase\cite{hep-ph/0207255}.

\subsection{Effect of electric charge neutrality}

In the asymmetric homogeneous quark matter under the electric charge neutrality,
u, d and s quarks have different densities, namely $\rho_d >\rho_u > \rho_s$,
or the  mismatched Fermi surfaces,
which disfavors the pairing. However,
when the system is heated,
the resulting smeared Fermi surfaces at finite $T$ make
the diquark pairing with mismatched Fermi surfaces possible, and hence
the diquark gap shows an abnormal temperature dependence that it has a maximum value
at a finite temperature\cite{hep-ph/0302142}.

Then, the competition between the chiral and the diquark
correlations becomes largest  at an intermediate  $T$, which
can lead to a nontrivial impact on the chiral-to-CSC transition\cite{hep-ph/0509172,arXiv:0808.3371}.
Indeed, Zhang, Fukushima and Kunihiro \cite{arXiv:0808.3371} showed
that there can appear
a crossover region in the phase boundary sandwiched by
novel two critical end points, although the resulting phase diagram is strongly dependent
on the choice of the parameters in the model Lagrangian.

\subsection{Combined effect of vector interaction and electric charge neutrality}

The combined effect of the vector interaction and the charge-neutrality condition
was examined by Zhang and Kunihiro\cite{arXiv:0904.1062}:
Although the structure of the phase diagram
depends on the choice of the strengths of the interaction,
there can appear three cross over region or two first-order critical lines,
 the ends of which are attached by critical points: the appearance of a crossover region
in the low $T$ region
including zero temperature is due to the vector coupling, whereas that in the intermediate $T$
region is understood to be due to the abnormal temperature dependence of the
diquark gap inherent for the asymmetric quark matter with mismatched Fermi surfaces.

\section{Incorporation of anomaly-induced chiral-diquark coupling}

Recently, Zhang and Kunihiro\cite{arXiv:1102.3263}
investigated the phase diagram
taking into account the following anomaly term
$\mathcal{L}_{\chi{d}}^{(6)}=K'/8\cdot\sum_{i,j,k=1}^3\sum_{\pm}$
$[({\psi}t_i^{f}t^{c}_k(1 \pm
\gamma_5){\psi}_C)(\bar{\psi}t_j^{f}t^{c}_k\\ (1
 \pm\gamma_5)\bar{\psi}_C)(\bar{\psi}_i( 1 \pm
\gamma_5)\psi_j)\label{eqn:Lagrangian4}]
$,
as well as the standard Kobayashi-Maskawa-'t Hooft term (with $K$ being the
strength) 
in addition to the four-Fermi scalar-pseudoscalar ($G_S$) 
 and the U(3)$_L\times$U(3)$_R$-invariant vector-axial-vector interactions.
Then the constituent quark masses $M_i$ and the dynamical Majarona masses $\Delta_i$
are expressed in terms of the chiral and diquark condensates,
$\sigma_i = \langle \bar\psi_i\psi_i \rangle$ and
$s_i =\langle \bar{\psi}_Ci\gamma_5 t_i^ft_i^c \psi \rangle$,
as follows($G_D$ being the diquark coupling):
$M_i =  m_i - 4G_S \sigma_i+K|\varepsilon_{ijk}|\sigma_j\sigma_k+ \frac{K'}{4}|s_i|^2$,
and
$
\Delta_i=2(G_D-\frac{K'}{4}\sigma_i)s_i
$,
respectively. 
 Here $m_u=m_d=5.5$ MeV and $m_s=140.7$ MeV denote the respective current quark masses.
The notable point is that $\mathcal{L}_{\chi{d}}^{(6)}$ induces the chiral-diquark
coupling as is manifested in the fourth and second term of $M_i$
and $\Delta_i$, respectively\cite{arXiv:1003.0408,arXiv:1007.5198}.

\begin{figure}
\hspace{-.0\textwidth}
\begin{minipage}[t]{.27\textwidth}
\includegraphics*[width=\textwidth]{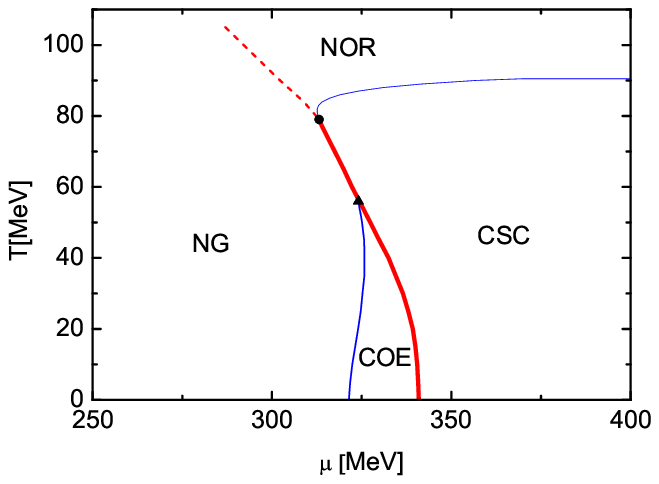}
\centerline{(a) $K'/K=2.0$}
\end{minipage}
\hspace{-.05\textwidth}
\begin{minipage}[t]{.27\textwidth}
\includegraphics*[width=\textwidth]{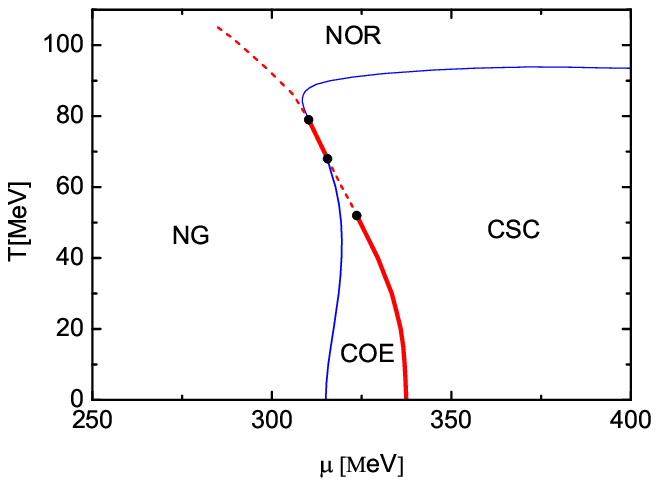}
\centerline{(b) $K'/K=2.25$}
\end{minipage}
\hspace{-.05\textwidth}
\begin{minipage}[t]{.27\textwidth}
\includegraphics*[width=\textwidth]{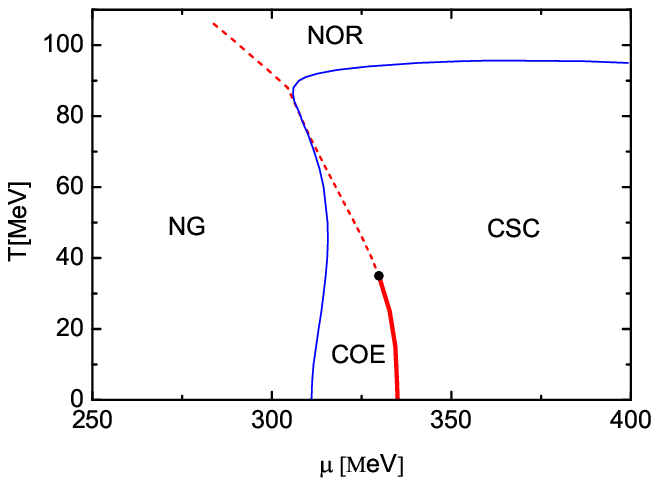}
\centerline{(c) $K'/K=2.4$}
\end{minipage}
\hspace{-.05\textwidth}
\begin{minipage}[t]{.27\textwidth}
\includegraphics*[width=\textwidth]{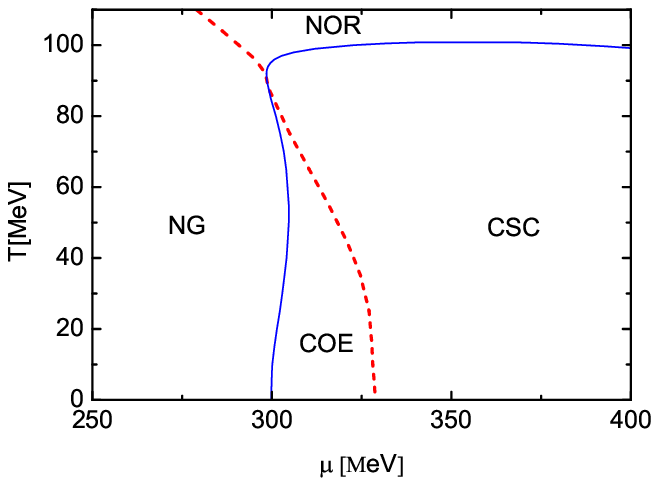}
\centerline{(d) $K'/K=2.8$}
\end{minipage}
\hspace{.0\textwidth}
 \caption{The phase diagrams
 for various values
of  $K'$ in the two-plus-one-flavor Nambu-Jona-Lasinio model with the charge-neutrality and
$\beta$-equilibrium being kept, but without the vector interaction.
 The thick-solid, thin-solid
and dashed lines denote the first-order, second-order
and chiral crossover critical lines, respectively. Adapted from
\citen{arXiv:1102.3263}.
}
\label{fig: pdzeroGv}
\end{figure}

\subsection{Without the vector term}

 In Fig.\ref{fig: pdzeroGv},
we first show the phase diagrams of the charge neutral quark matter for
various ratio $K'/K$ when the vector term is absent ($G_V=0$)\cite{arXiv:1102.3263}:
When the ratio $K'/K$ is as small as 2.0, we have the standard phase diagram with a
single critical point, although the existence of the combined chiral-diquark phase
in the low-$T$ region is notable. Then $\kr$ is increased up to 2.2,
the crossover window opens in the intermediate-$T$ region, inherent for the
charge-neutral system with mismatched Fermi surfaces;
 we note that the
competition between the chiral and diquark correlations  is enhanced by
the new anomaly term. When $K'$ is further increased, the diquark correlation
becomes so large that the chiral transition in the CSC phase
becomes smooth, and eventually the first-order critical line disappears completely
in the phase diagram. Notice, however, that the crossover window never opens in
the low-$T$ region including zero temperature, as is the case without the
charge neutrality constraint\cite{arXiv:1007.5198}, which is contrary to the
observation in \citen{arXiv:1003.0408}.

\begin{figure}
\hspace{-.0\textwidth}
\begin{minipage}[t]{.27\textwidth}
\includegraphics*[width=\textwidth]{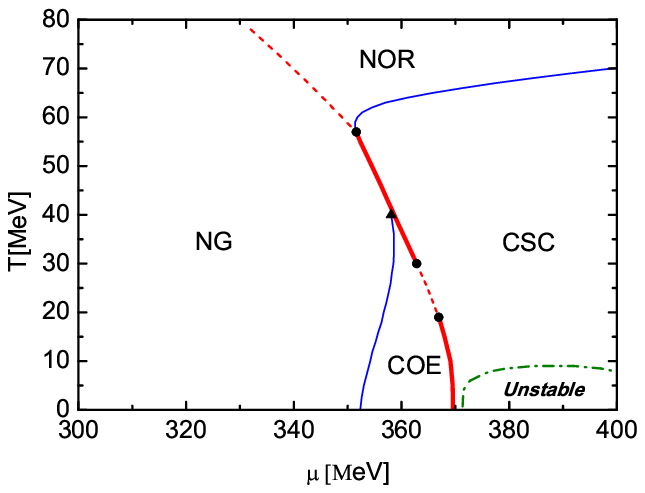}
\centerline{(a) $K'/K=0.55$}
\end{minipage}
\hspace{-.05\textwidth}
\begin{minipage}[t]{.27\textwidth}
\includegraphics*[width=\textwidth]{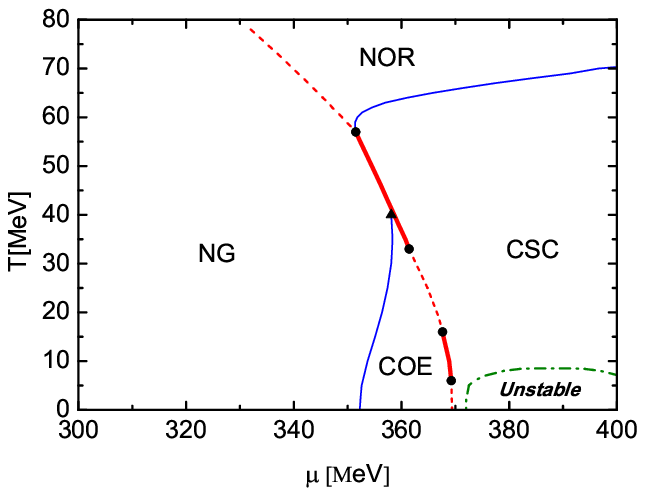}
\centerline{(b) $K'/K=0.57$}
\end{minipage}
\hspace{-.05\textwidth}
\begin{minipage}[t]{.27\textwidth}
\includegraphics*[width=\textwidth]{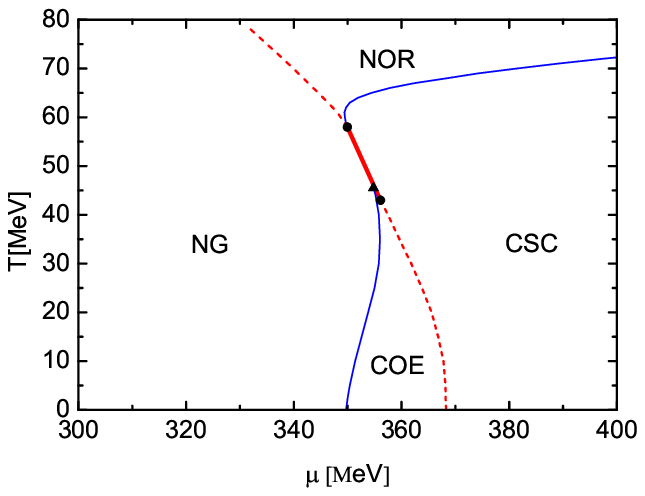}
\centerline{(c) $K'/K=0.70$}
\end{minipage}
\hspace{-.05\textwidth}
\begin{minipage}[t]{.27\textwidth}
\includegraphics*[width=\textwidth]{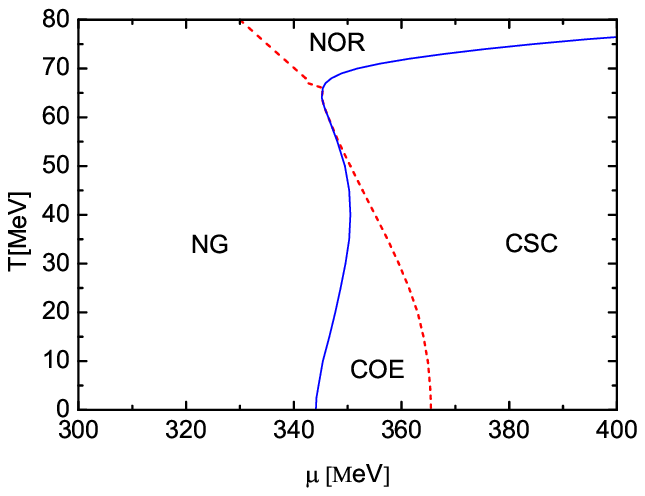}
\centerline{(d) $K'/K=1.0$}
\end{minipage}
\caption{The phase diagrams of charge-neutral quark matter
for several values of $K'/K$ and fixed $G_V/G_S=0.25$.
Adapted from \citen{arXiv:1102.3263}.}
\label{fig:pdGVfixed}
\end{figure}

\subsection{Combined effect of vector and anomaly terms}

We show in Fig.\ref{fig:pdGVfixed} the phase diagrams
when the vector term is also included as well as the anomaly terms under
charge-neutrality constraint\cite{arXiv:1102.3263}.
Again there appears a crossover region in the intermediate $T$ region
but for much smaller $\kr$ this time.
Slightly increasing $\kr$ up to 0.57,
a new crossover window opens with a new critical point attached
 in the low-$T$ region including zero temperature,
which is again due to the chiral-diquark competition enhanced by
the vector term as well as the anomaly term.
As $\kr$ is increased further, the island of the first-order critical line
for the chiral restoration disappears owing to so strong diquark correlation, and
the first-order critical line ceases to exist for a realistic value $\kr=1.0$.
We  note that the unstable region characterized
by the chromomagnetic instability (bordered by the dash dotted line)
tends to shrink and ultimately vanishes in the phase diagram\cite{arXiv:1102.3263}.

\section{Summary}

The diquark-chiral coupling induced by $U(1)_A$ anomaly plays
a similar role as the vector interaction for the phase diagram
of the quark matter under charge-neutrality constraint.
As a result, the phase boundary for the chiral-color-superconducting
(CSC)  phase transition
can have alternate multiple windows of the crossover/first-order transition lines attached
with critical point(s). The message to be taken from the present
mean-field level calculation 
is that the QCD matter under the charge neutrality constraint
is soft for the simultaneous formation of chiral and diquark condensates
around the would-be phase boundary, implying possible absence of the first-order
transition line and critical points.
Since the chiral transition at finite
density involves a change of baryon density, the soft mode
is actually a combined fluctuations of  chiral, diquark and baryon densities\cite{Kunihiro:2010vh}.

T. K. was supported by the Grant-in-Aid for the Global COE Program
``The Next Generation of Physics, Spun from Universality and Emergence'' 
and also by a Grant-in-Aid for Scientific Research (No.20540265, 2334006)  from
MEXT of Japan.
Z. Z. was supported by the Fundamental Research Funds
for the Central Universities of China. 

%

\end{document}